\documentclass[article]{sa}

\author{Francesco Chiodo\\University Magna Graecia of Catanzaro \And 
         Pietro Hiram Guzzi\\Magna Graecia University of Catanzaro}
\title{Towards a Recommender System for Profiling Users in a Renewable Energetic Community.}

\Plainauthor{Georg Simmel, Second Author} 
\Plaintitle{A Capitalized Title: Something about a Package for} 
\Shorttitle{A Capitalized Title} 

\Abstract{

Current Energy systems located in almost all nations are going through a radical transformation motivated by technological, environmental and institutional needs. The main factors leading this transformation are the introduction of novel technologies for energy production and storing, the insurgence of climate change, and the attention to the introduction of low-impact technologies in some countries. Here we focus in particular on introducing relatively small  community energy systems based on solar energy that aim  to re-organize local energy systems to integrate distributed energy resources and engage local communities. Each community has a set of producers and a set of consumers (and a set of producers/consumers called prosumers). One of the key aspects of the energetic communities is to maximise the energy that is shared within the user. Thus, it is crucial to select the best consumers/prosumers on the basis of their profile of consumption, in order to minimize subsequent management of the energy once the community is built. Here we describe the design of a recommender sysstem that can profile user on the basis of their past profile for the subsequent admission into the energetic community. \\ \textbf{Experiments supporting this publication have been carried under the BDTI (Big Data Test Infrastructure) of the European Union. The contents of this publication are the sole responsibility of authors and do not necessarily reflect the opinion of the European Union.}
}
\Plainkeywords{keywords, comma-separated, not capitalized} 

\Journal{Preprint}
\Volume{}
\Issue{} 
\Pages{} 
\Year{}
\Submitdate{\today} 
\Acceptdate{\today} 
\DOI{URL/DOI GOES HERE} 

\Address{
  Dott. Francesco Chiodo\\
  University Magna Graecia\\
  Catanzaro, Italy\\
  E-mail: \email{francesco.chiodo.op@gmail.com}\\
}

\begin{document}


\section{Introduction}

Consumers have been for many years the last step of the production-transport-use chain of energy without the possibility of the kind of production (nuclear, fossil or renewable production systems). Recently, the evolution of the laws of the EU governments has placed users at the centre of renewable energy projects, by creating novel models of energy systems. In particular, the policy of the EU is to create a transition towards the creation of \textit{fully active actors}, i.e users that are both consumers and producers, i.e. prosumers, of renewable energy. The main consequence of this policy is that people are fully involved in  the production, distribution, storage, and end use of energy. This change may be achieved through the creation and development of Renewable Community Energy (RCE).

The phrase \textit{renewable energy community} is used to describe many similar projects whose common element is the union of many little (less than 200 kw depending on the country), producers of energy from renewable resources with many little consumer (usually families or small enterprises) forming a relatively small subgrid (or community) of energy transportation. The former definition admits many slightly different meanings due to different legal and economic models abound and that depending on the local context.

One of the key problems in a renewable energetic community is the maximisation of the energy that is shared among the user, i.e. the amount of the energy that is consumed by the users with respect to the produced energy. 



\begin{figure}
    \centering
    \includegraphics[width=0.7\textwidth]{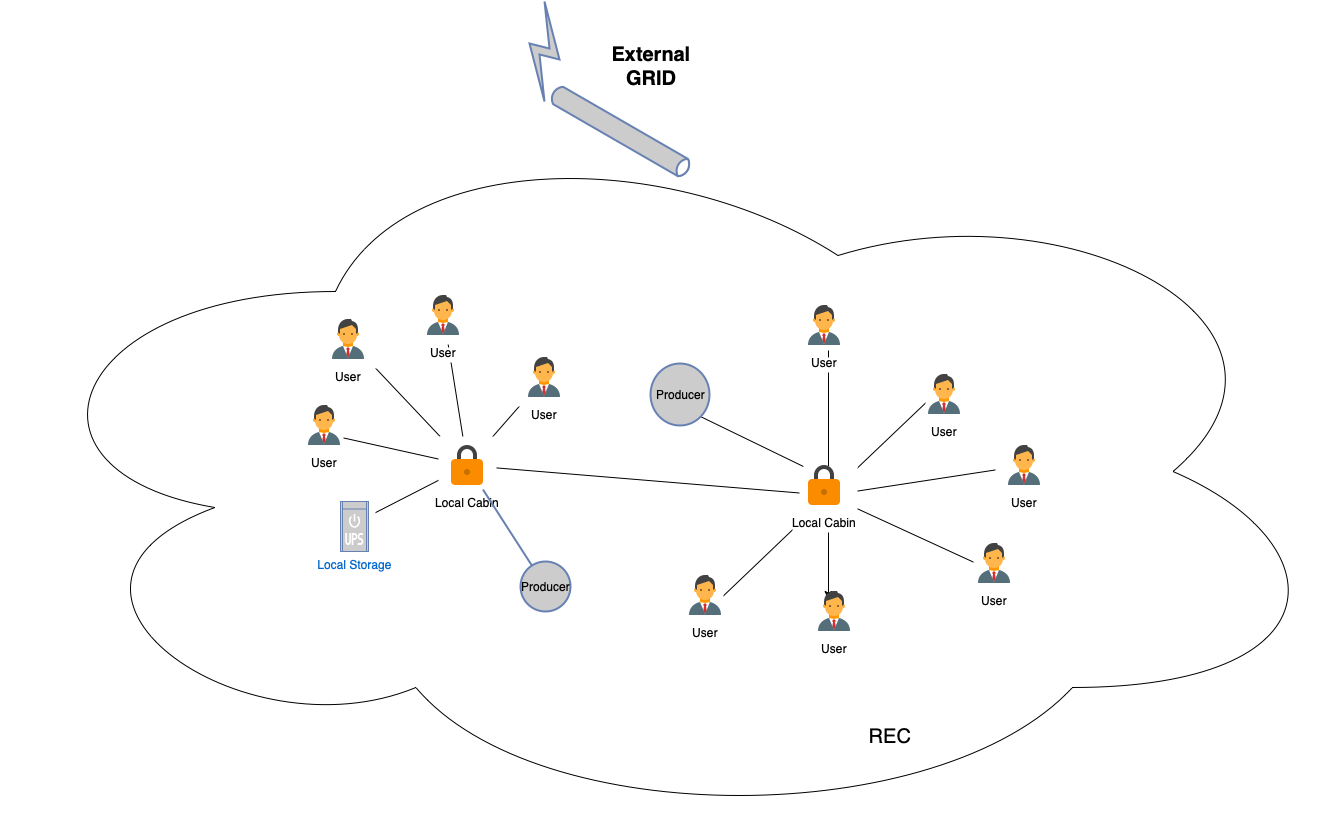}
    \caption{Renewable Energetic Community in a Grid Context.}
    \label{fig:REC}
\end{figure}

Actors of a renewable energetic community fall into two major classes: (i) {\bf producers}, i.e. user that push into the REC energy, (ii) {\bf consumers}. We know that a user can simultaneously be producer and consumer, so for lack of simplicity we split this user into two dummy users. 
We focus, in particular, on energetic communities based on solar energy. Figure \ref{fig:REC} depicts an example of our model.

Consequently, there is the need to admit  (or recommend to admit) into the energetic community user whose profile of the consumption of the energy has the best fit to the profile of the producers. We model such a problem as a recommender system \cite{agapito2016dietos,agapito2019parallel}. The production of the energy is not directly controllable by the user since it depends on the weather, and we may supply only with local storage system, we have some main parameters: (i) the maximum production, (ii) an average profile of production, (iii) the maximum storage capacity. Therefore, we want to profile users on the basis of their past consumption profile and then recommend the introduction of the admission into the REC. This is similar to the design of a recommender system through  collaborative filtering \cite{schafer2007collaborative}. 

The idea is that people that had a similar profile of consumption in the past will have a similar profile in the future. The system generates recommendations using only information about rating profiles for different users or items. The main difference of our approach  with respect  classical collaborative filtering is that, after profiling users, we admit a transductive capacity, to rate the user based on the other users present into the energetic community, given that we have an ideal model of users presents into the community. In this way, after that user have been profiled, the REC decisor may suggest to accept or not the novel user into the REC.


\begin{figure}
    \centering
    \includegraphics[width=0.8\textwidth]{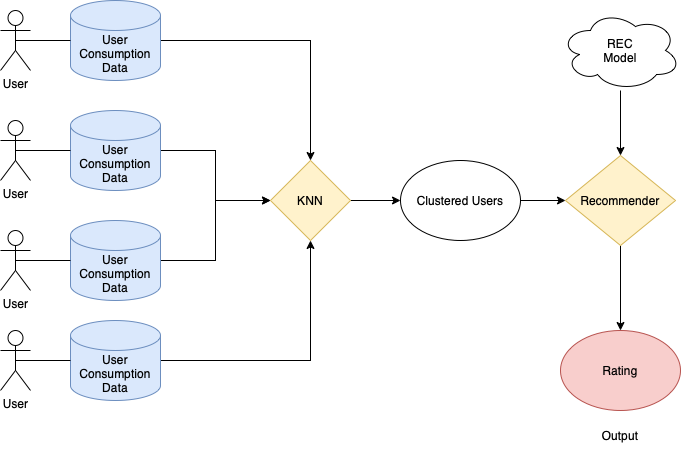}
    \caption{The Proposed Framework.}
    \label{fig:recommender}
\end{figure}

\section{Methods}

In this collaboration with the {\bf BDTI} (Big Data Technology Infrastructure) the use of an unsupervised machine learning algorithm make possible the profiling of the energetic users in a renewable energetic community.\\There are 3 main types of profile: families without children, families with children and commercial activities. For each type of profile, there are some examples, for a total of 10 monitoring examples for 2 months.\\ After archiving all the data inside an SQL database, the work was carried out on 10 CSV files, one for each profile. In this CSV files, there were the consumption data in function of time, divided in moths, days and hours.\\ By testing supervised and unsupervised machine learning algorithms, the best result (calculated using the ROC curve and other evaluation metrics) was obtained with K-means algorithm with 3 clusters.\\This algorithm make a partition of data in a predefined number of clusters and it stops to iterate when the within-cluster variation is as small as possible.\\The choice to use Python as programming language for the development of this application is for the libraries for CSV data management and the libraries that let an easy build of Machine Learning algorithms.

\subsection{Datasets.}
Data are archived in CSV files that represent the hour's consumption for 12 months, taking into account the passing of moths and seasons, with the different consumption for different periods.\\
This CSV have different fields: year (for implementing new data for making more precise our clustering algorithm), month, day, hour and energy consumption in kW/h, making possible an easy an intuitive data analysis (consumption/time) and visualization through graphs.\\

\subsection{Algorithms.}
The most efficient algorithm to process this data turned out to be the K-MEANS algorithm.\\It is an unsupervised machine learning algorithm based on the concept of centroids, it computes and iterates the positions of these centroids until it founds the optimal solutions.\\In the name of the algorithm, the letter "K" means the number of clusters, that we have to decide before starting the iterations.\\In the first step we place K centroids in random places and start to iterate.\\In every iteration calculates the euclidean distance between each element and each center, assign each element to the nearest cluster center (centroids) and recalculate the centers for each cluster.\\While the position of the centroids continue to change, the algorithm continue to iterate.\\ When the position of all centroids stop to move or they move in a pre-defined tolerance value, the algorithm go out and give the output.\\ Using new data, when we succeed in recording them,  make the algorithm stonger and more precise.\\

\begin{figure}
    \centering
    \includegraphics[width=0.8\textwidth]{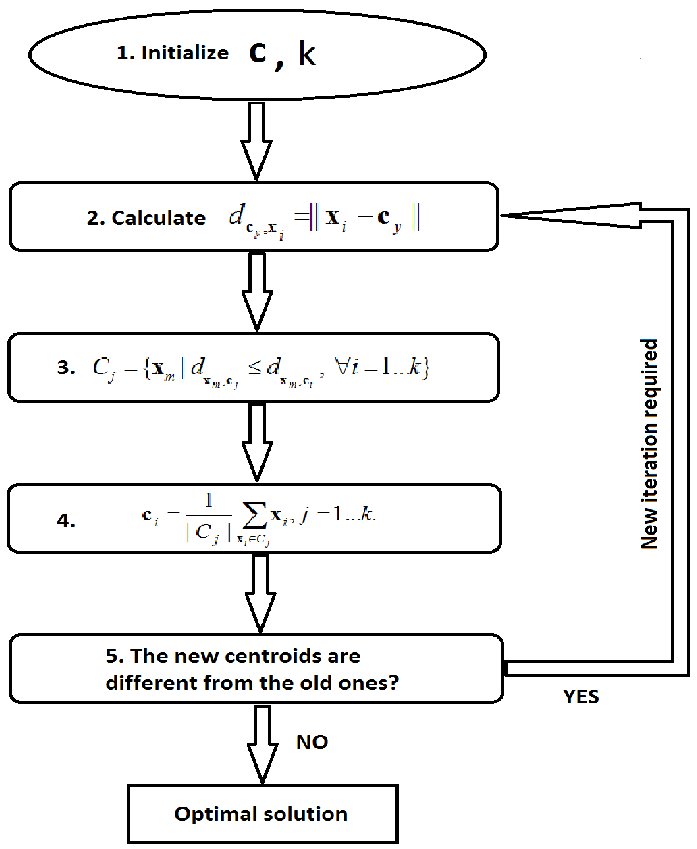}
    \caption{The Proposed Framework.}
    \label{fig:k-step}
\end{figure}



\section{Related Work}

\subsection{Recommender Systems}

A recommender system is defined as a kind of information filtering system that receives as input an user and set of items and predicts the rate (or the preference) that the user gives to the items \cite{bobadilla2013recommender}. Such systems are commonly used to suggest some items to a consumer in a on line marketplace, or in video and music services and in open web content recommenders. 

According to ACM, recommender systems are a subclass of information systems that are focused on information retrieval task. This general definition needs to be adapted in the many applications in different scenarios in which recommender systems have been customised to achieve specific goals. Literature reports first applications of Recommenders Systems for content personalisation of on line systems on the basis of user preferences. In parallel RS have been also applied  to energy management \cite{alsalemi,sardianos2019want,AGUILAR2021111530} with the goal to reshape energy profiles.

\begin{figure}[ht]
    \centering
    \includegraphics[width=0.8\textwidth]{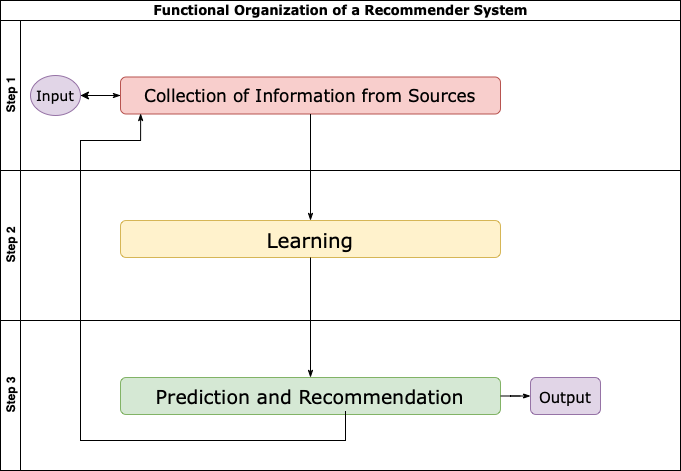}
    \caption{Functional Organisation of a Recommender System}
    \label{fig:functionalorganization}
\end{figure}

It should be noted RS maintain a shared logic of approach despite of the high difference of application scenarios in which they are applied \cite{HIMEUR20211} as depicted in Figure \ref{fig:functionalorganization}. 

In the first step, RS has to collect all the needed information about the user integrating when needed many information sources (e.g. databases, crawler or through sensors). In this step it is crucial to maintain the highest data quality possible since it may seriously impact the subsequent steps. In the second step the system first extract most relevant features and then it trans the most suitable model to analyse relationship among the users and the “items”. Such relationships are used to create recommendation in the third step.
In the third step, the system predicts the user-to-items preferences using the pre-trained model and ranks the items most likely to fit the  preferences of the user(s).

\section*{Acknowledgement}

{\bf Experiments supporting this publication have been carried under the BDTI (Big Data Test Infrastructure) of the European Union. The contents of this publication are the sole responsibility of authors and do not necessarily reflect the opinion of the European Union.}

\bibliography{sa}

\end{document}